\documentclass[12pt]{article}
\usepackage{amsfonts,amssymb}
\usepackage{mathrsfs,color}
\usepackage{graphicx}
\usepackage{subfig}

\newcommand{\DD}{{\mathscr{D}}}
\newcommand{\R}{{\mathbb{R}}}

\newcommand{\C}{{\mathbb{C}}}

\newcommand{\CP}{{\mathbb{C}}{{P}}}

\newcommand{\ee}{\mathcal{E}}
\newcommand{\beq}{\begin{equation}}
\newcommand{\eeq}{\end{equation}}
\newcommand{\bea}{\begin{eqnarray}}
\newcommand{\eea}{\end{eqnarray}}
\newcommand{\ra}{\rightarrow}

\newcommand{\cd}{\partial}

\newcommand{\ip}[1]{\langle#1\rangle}

\newcommand{\ignore}[1]{}

\def \d{\mathrm{d}}

\newcommand{\ol}{\overline}

\renewcommand{\phi}{\varphi}

\begin{document}

\title{Supercurrent coupling destabilizes knot solitons}
\author{
J. J\"aykk\"a\thanks{E-mail: {\tt juhaj@iki.fi}}\,\, and 
J.M. Speight\thanks{E-mail: {\tt speight@maths.leeds.ac.uk}}\\
School of Mathematics, University of Leeds\\
Leeds LS2 9JT, England
}

\date{}
\maketitle

\begin{abstract}
In an influential paper of 2002, Babaev, Faddeev and Niemi conjectured
that two-component Ginzburg-Landau (TCGL) theory in three dimensions 
should support knot solitons,
where the projective equivalence class of the pair of complex condensate
fields $[\psi_1,\psi_2]:\R^3\ra \CP^1$ has non-zero Hopf degree. The
conjecture was motivated by a certain truncation of the TCGL model
which reduced it to the Faddeev-Skyrme model, long known to
support knot solitons. Physically, the truncation amounts to ignoring the
coupling between $[\psi_1,\psi_2]$ and the supercurrent of the condensates.
The current paper presents a direct test of the validity of this truncation by
numerically tracking the knot solitons as the supercurrent coupling is turned 
back on. It is found that the knot solitons shrink and disappear as the
true TCGL model is reached. This undermines the reasoning underlying the
conjecture and, when combined with other negative numerical studies,
suggests the conjecture, in its original form, is very unlikely to be true.
\vspace*{0.3cm}\newline
\end{abstract}

\maketitle

\section{Introduction}\label{intro}

Two component Ginzburg-Landau (TCGL) theory is a phenomenological field theory
which is enjoying increasing prominence in several areas of condensed matter
theory. It consists of two complex scalar fields $\psi_1,\psi_2:\R^3\ra\C$
minimally coupled to a $U(1)$ gauge field $A\in\Omega^1(\R^3)$ and (perhaps)
each other via an energy density of the form
\beq
\ee=\frac12|\d_A\psi_1|^2+\frac12|\d_A\psi_2|^2+\frac12|\d A|^2+U(\psi_1,\psi_2)
\eeq
where $\d_A\psi_a=\d\psi_a-iA\psi_a$ and $U$ is a potential function
whose details depend strongly on the precise physical context.
In an influential and much-cited
paper of 2002 \cite{babfadnie}, Babaev, Faddeev and
Niemi conjectured that models of this type should possess ``knot solitons'' in 
which the projective equivalence class of the pair of complex fields
$\psi_1,\psi_2$,
\beq
\phi=[\psi_1,\psi_2]:\R^3\ra\CP^1\cong S^2,
\eeq
has non-zero Hopf degree. This conjecture is based on the observation that,
when expressed in terms of the gauge-invariant fields $\phi$,
$\rho=\sqrt{|\psi_1|^2+|\psi_2|^2}$ and
\beq\label{C}
C=\frac{i}{2\rho^2}\sum_{a=1}^2(\ol{\psi}_a\d_A\psi_a-\psi_a\ol{\d_A\psi_a})
\eeq
the TCGL energy density is
\beq
\ee=\frac18\rho^2|\d\phi|^2+\frac12|\d C+\frac12\phi^*\omega|^2
+\frac12|\d\rho|^2+\frac12\rho^2|C|^2+U(\rho,\phi),
\eeq
where $\omega$ is the area form of the usual metric on the unit two-sphere.
In superconductivity contexts, $\rho^2 C$ is the total supercurrent associated
with condensates $\psi_1,\psi_2$ and so, in a slight abuse of terminology,
we shall refer to $C$ itself as the supercurrent.
If we impose that $\rho=\rho_0>0$, a constant, and $C\equiv 0$, $\ee$
coincides with the energy density of the Faddeev-Skyrme (FS)
model, which is
known to possess knot solitons of every Hopf degree $Q$. These are smooth,
spatially localized
global energy minimizers in their homotopy class. 
Topologically, maps  $\phi:\R^3\cup\{\infty\}\ra S^2$ are
classified by the framed cobordism class of $\phi^{-1}(p)$ where $p\in S^2$
is any regular value of $\phi$. In general, $\phi^{-1}(p)$ is a union of closed
curves in $\R^3$, that is, a union of knots. In the case of FS solitons
these knots are simple unknots for low $Q$, but become more complicated
as $Q$ grows.
For a survey of FS knot solitons, see \cite{ratvol}. 

So the conjecture of \cite{babfadnie} results from identifying the FS model
as a ``submodel'' of the original TCGL model. In replacing the TCGL model by 
its FS submodel, one is making a two-step
truncation. The first truncation,
imposing $\rho=\rho_0>0$ constant, is reasonable for suitable choices of $U$,
for example
\beq\label{U}
U=\lambda(\rho_0^2-|\psi_1|^2-|\psi_2|^2)^2
\eeq
in the limit of large $\lambda$. Indeed the analogous truncation
in {\em ungauged} two-condensate systems has been found to introduce no
drastic qualitative changes \cite{batcoosut}.
However, the second truncation, imposing
$C=0$, is harder to justify since, even for $\rho$ constant, $C$ is coupled
to $\phi$. Indeed, Babaev, Faddeev and Niemi do not directly claim that 
setting $C=0$ is a good approximation for the TCGL model \cite{babfadnie}.
Nonetheless, their claim that TCGL theory should support knot solitons assumes
that the coupling between $\phi$ and $C$ does not destroy the solitons, and
hence, that the second truncation $C=0$ is, if not a ``good approximation'',
valid at least at a qualitative level. This is an assumption, not
an established fact, a distinction missed by the authors
of many of the papers citing \cite{babfadnie}, who
treat the existence of knot solitons in TCGL models likewise
as established fact rather
than conjecture.

There have been many attempts to construct knot solitons in TCGL models 
numerically, by choosing $\psi_1,\psi_2,A$ with $\phi=[\psi_1,\psi_2]$
having $Q>0$, usually in some toroidally symmetric ansatz,
 and minimizing $E=\int_{\R^3}\ee$ using some
gradient descent method. In almost all cases the initial field
configurations have shrunk and fallen through the lattice mesh
\cite{jay2,jayhiesal,war}. 
The only exception we know of is \cite{niepalvir}, which uses direct
gradient flow to evolve the field configuration towards a critical
point of $E$, and claims to have found ``definite convergence towards
torus-shaped configurations''. Given that this finding has not been reproduced
by any other researchers, it seems likely that this is an artifact of
having used too lax a convergence criterion (unfortunately, the precise
convergence criterion used is not described in \cite{niepalvir},
but gradient flow is a notoriously slow method, and it is easy to mistake
a slowly evolving field for a truly static one).

The failure of the direct approach to find knot solitons is not surprising given
the results of \cite{spe}. Even if one imposes that $\rho$ is
non-vanishing, so that $Q$ is well-defined, in every degree class the
infimum of $E(\psi_1,\psi_2,A)$ is $0$ because, for
suitably chosen $A$, any spatially localized configuration $\psi_1,\psi_2$
is unstable against Derrick scaling. This shows that, for all $Q\neq 0$, there
is no {\em global} energy minimizer. 
Note that this contradicts the common misconception, repeated in 
\cite{niepalvir}, that the existence of a nontrivial topological charge
protects a field configuration from collapse. 
So knot solitons, if they
exist in TCGL theory, can only be {\em local} minimizers of $E$, 
and
finding them is like looking for a hollow dip in a hill-top, rather than
the floor of a valley: unless one's initial guess is very close to the 
 dip, one is very unlikely to find it. 
Indeed this point was understood, from a more qualitative, physical
viewpoint, by Babaev \cite{bab}, some years after making the original
conjecture.
For this reason, the failure
of previous numerical studies to find knot solitons, while discouraging,
does not systematically undermine the conjecture of \cite{babfadnie}. To
do so using the direct approach one would have to systematically search
the infinite dimensional space of all possible initial data.

In this paper we pursue a different approach. We study the TCGL model in the
case of a hard-confining potential, (\ref{U}) with $\lambda\ra\infty$, or,
equivalently, in the sigma model limit, where 
$|\psi_1|^2+|\psi_2|^2=\rho_0^2$. We may, without loss of generality, take
$\rho_0=1$.  This allows us to concentrate on the dynamically crucial
issue of the coupling between $\phi$ and $C$. Rather than minimizing
$E$ starting with some plausible initial guess, we consider the one-parameter
family of models with energies
\beq\label{Ealpha}
E^{(\alpha)}=\frac18\|\d\phi\|^2+\frac18\|\phi^*\omega\|^2+
\frac\alpha2\ip{\d C,\phi^*\omega}+\frac12\|\d C\|^2
+\frac12\|C\|^2
\eeq
parametrized by $\alpha\in[0,1]$, where $\ip{\cdot,\cdot}$ denotes 
$L^2$ inner product and $\|\cdot\|$ denotes $L^2$ norm. Note that at
$\alpha=1$ this is the TCGL model (in the sigma model limit) while at 
$\alpha=0$ it decouples into the FS model and an uncoupled Proca model
for $C$. So the $\alpha=0$ model clearly supports knot solitons with 
$C\equiv 0$ - they are simply the usual FS knot solitons. The question is then,
what is the fate of these solitons as the supercurrent coupling is 
``turned on'', that is, $\alpha$ is increased to $1$?
We address this problem numerically by a continuation method. Starting at
$\alpha=0$, we numerically minimize $E^{(\alpha)}$ within a given homotopy class
using 
the limited memory quasi-Newton algorithm (also called a variable
metric algorithm) with BFGS  formula for
Hessian approximations \cite{numrec}.
Having found a minimizer for this
value of $\alpha$, we increase $\alpha$ slightly and minimize $E$
again, starting from the energy minimizer just obtained. In this way
we construct a curve of energy minimizers parametrized by $\alpha$, and the
crucial question is whether this curve continues all the way to $\alpha=1$.

We will see that, for all $\alpha\in[0,1)$, we have a lower energy bound
\beq
E^{(\alpha)}\geq  c_0\sqrt{1-\alpha}|Q|^{\frac34}
\eeq
where $c_0>0$ is an absolute constant. This leads one to expect the
knot solitons to persist whilever $\alpha<1$, albeit with nonzero supercurrent
$C$, and
indeed this is what we find. We shall produce convincing
numerical evidence, however, that as $\alpha\ra 1$, these knot solitons
shrink to zero size, because $C$ deforms precisely to the form for
which $E^{(1)}$ is not stable against Derrick scaling.

Note that our results are not just another negative finding in the same vein as
\cite{jayhiesal}, since our approach tests not just the conjecture of \cite{babfadnie},
but also the reasoning underlying it. That is, our results conclusively
demonstrate that, notwithstanding the formal similarity between the
TCGL model (in suitable variables) and the FS model, the knot solitons of the
latter definitely do {\em not} persist in the former, even in the most
favourable case where a hard confining potential enforces 
$|\psi_1|^2+|\psi_2|^2=1$. This also elucidates the mechanism by
which knot solitons are destabilized. It has nothing, in general, to do
with $\rho$ shrinking and aquiring isolated zeroes (thus allowing $Q$ to
discontinuously drop). Even in the sigma model limit, where $Q$ is
a rigorously defined topological invariant, the coupling to the 
supercurrent $C$ alone suffices to destabilize the knot solitons.
Of course, this does not show for certain that $E$ has no local minima
with $Q\neq 0$, but it does invalidate the reasoning used to motivate the
conjecture in the first place. The present results complement work
of Ward \cite{war} (extended by one of us \cite{jay2}) which also 
embeds TCGL theory into a one-parameter family of models one end of which
supports solitons. Again, these solitons disappear as the true TCGL
theory is approached. It also complements the exact results of
\cite{spe} where it was shown analytically that supercurrent coupling 
destabilizes the $Q=1$ Hopf soliton on physical space $S^3$. 
In the absence of evidence to the contrary, one should wield Occam's razor and
conclude that, in all likelihood, the basic TCGL theory does not possess knot 
solitons. There remains the possibility that more elaborate,
but physically relevant, versions of
TCGL theory, involving direct current-current interactions, may have knot
solitons \cite{bab}, but until direct evidence for this is found, 
we prefer to remain sceptical.

\section{Energy bound and Derrick scaling}\label{ebds}

We seek local minimizers of the energy functional $E^{(\alpha)}$ defined in (\ref{Ealpha}) with $\phi:\R^3\ra S^2$ and
$C\in\Omega^1(\R^3)$ having the boundary behaviour
\beq\label{bc}
\lim_{|x|\ra\infty}\phi(x)=\phi_\infty,\qquad
\lim_{|x|\ra\infty}C(x)=0
\eeq
and $\alpha\in[0,1]$. Without loss of generality, we choose 
$\phi_\infty=(0,0,1)$. Configurations with this boundary behaviour fall into
disjoint homotopy classes labelled by the integer Hopf invariant
\beq
Q=\frac{1}{16\pi^2}\int_{\R^3}a\wedge \d a
\eeq
where $a$ is any one-form on $\R^3$ such that $\d a=\phi^*\omega$. It is
well known that the Faddeev-Skyrme energy, which coincides (up to a factor of
$\frac14$ in the usual normalization) with $E^{(0)}(\phi,0)$ satisfies a
topological lower energy bound
\beq
\label{VK}
E^{(0)}(\phi,0)\geq c_0|Q(\phi)|^\frac34
\eeq
where $c_0>0$ is an absolute constant \cite{vakkap}. Now, for all 
$\alpha\in[0,1]$,
\bea
E^{(\alpha)}&=&\frac18\|\d\phi\|^2+\frac18(1-\alpha)\|\phi^*\omega\|^2
+\frac12(1-\alpha)\|\d C\|^2+\frac\alpha2\|\d C+\frac12\phi^*\omega\|^2+
\frac12\|C\|^2\nonumber\\
&\geq&\frac18\|\d\phi\|^2+\frac18(1-\alpha)\|\phi^*\omega\|^2\nonumber\\
&=&\frac18\sqrt{1-\alpha}\left(\|\d\hat\phi\|^2+\|\hat\phi^*\omega\|^2\right)
\eea
where $\hat\phi(x)=\phi(\sqrt{1-\alpha}x)$. Hence, by the usual
Vakulenko-Kapitanski bound (\ref{VK}),
\beq\label{ebound}
E^{(\alpha)}(\phi,C)\geq c_0\sqrt{1-\alpha}|Q(\phi)|^\frac34.
\eeq
This leads us to expect that $E^{(\alpha)}$ will have a global energy minimizer
in each homotopy class whilever $\alpha\in[0,1)$. 

It is instructive to subject $E^{(\alpha)}$ to the Derrick scaling test
for
all $\alpha\in[0,1]$ since this gives an integral constraint on $(\phi,C)$
which we can use as a consistency check on our numerical scheme \cite{der}. 
Assume
that $(\phi,C)$ is a critical point of $E^{(\alpha)}$. Then
$E^{(\alpha)}$ is stationary with respect to all variations of
$(\phi,C)$ and so, in particular, with respect to the variation
\beq\label{jgibd}
\phi_\lambda=\phi\circ\DD_\lambda,\qquad
C_\lambda=\DD_\lambda^* C
\eeq
where $\lambda\in(0,\infty)$ and $\DD_\lambda:\R^3\ra\R^3$ is the dilation map
$\DD_\lambda(x)=\lambda x$. Note that $\phi_1=\phi$ and $C_1=C$. A short
calculation shows that
\beq\label{elambda}
E(\lambda):=E^{(\alpha)}(\phi_\lambda,C_\lambda)
=\frac{1}{8\lambda}\|\d\phi\|^2+\frac\lambda8\|\phi^*\omega\|^2
+\frac{\lambda\alpha}{2}\ip{\phi^*\omega,\d C}+\frac\lambda2\|\d C\|^2
+\frac{1}{2\lambda}\|C\|^2.
\eeq
Hence, since $(\phi,C)$ is a critical point, 
\beq\label{DC1}
E'(1)=-\frac{1}{8}\|\d\phi\|^2+\frac18\|\phi^*\omega\|^2
+\frac{\alpha}{2}\ip{\phi^*\omega,\d C}+\frac12\|\d C\|^2
-\frac{1}{2}\|C\|^2=0.
\eeq
We shall refer to this as the Derrick constraint. 
\ignore{It is straightforward
to derive another integral constraint in similar fashion, by considering the 
variation
\beq
\phi_\mu=\phi,\qquad C_\mu=\mu C
\eeq
where $\mu\in(0,\infty)$. Again, if $(\phi,C)$ is a critical point of
$E^{(\alpha)}$ then $d E^{(\alpha)}(\phi_\mu,C_\mu)/d\mu=0$ at $\mu=1$, whence
\beq\label{DC2}
\frac\alpha2\ip{\phi^*\omega,\d C}+\|\d C\|^2+\|C\|^2=0,
\eeq
which we shall refer to as the second Derrick constraint.} 

Note that, in the case $\alpha=1$ (the TCGL model), if $C$ is
chosen so that 
\beq\label{nasty}
\d C+\frac12\phi^*\omega=0,
\eeq
and $(\phi_\lambda,C_\lambda)$ defined as in (\ref{jgibd}) we have
\beq
E^{(1)}(\phi_\lambda,C_\lambda)=\frac{1}{8\lambda}\|\d\phi\|^2+\frac{1}{2\lambda}\|C\|^2\ra0
\eeq
as $\lambda\ra\infty$. For all $\phi$ satisfying (\ref{bc}) there exists
$C$ satisfying (\ref{nasty}) and (\ref{bc}), which shows that
$E^{(1)}(\phi,C)$ has infimum $0$ in every homotopy class \cite{spe}.
This argument, which
 is essentially equivalent to the one described in \cite{ratvol}
(whose authors attribute it to unpublished work of Forgacs and Volkov),
explains why the energy bound (\ref{ebound})
becomes trivial at $\alpha=1$.
Of course, it does {\em not} follow that $E^{(1)}$ can have no critical points,
because there is no reason why critical points of $E^{(1)}$ should satisfy
(\ref{nasty}).

For the purposes of numerics, we will study $E^{(\alpha)}$ not
on $\R^3$, but in a large box $B=[-L,L]\times[-L,L]\times[-L,L]$ with
Dirichlet boundary conditions $\phi(x)=\phi_\infty$, $C(x)=0$, for
all $x\in\cd B$. 
\ignore{In this case, the second scaling argument above
proceeds without change, so any minimizer for the box $B$ should satisfy
(\ref{DC2}).}
 In this case, 
the Derrick scaling argument must be modified, as follows.
First, we note that the variation $(\phi_\lambda, C_\lambda)$ is only well
defined for $\lambda\in[1,\infty)$, since for $\lambda\in(0,1)$ $(\phi_\lambda,
C_\lambda)$ does not satisfy the boundary conditions. For $\lambda\geq 1$
we first notionally extend $\phi, C$ to the whole of $\R^3$
by their boundary values, then define $(\phi_\lambda, C_\lambda)$
as in (\ref{jgibd}). The equation (\ref{elambda}) for $E(\lambda)$
still holds, but now on $[1,\infty)$ rather than $(0,\infty)$. We
require that $E(\lambda)$ has a local minimum at $\lambda=1$, but now,
since this is at an end-point of the domain, it does {\em not} follow that 
$E'(1)=0$. Rather, we know only that
\beq\label{DC1B}
E'(1)=-\frac{1}{8}\|\d\phi\|^2+\frac18\|\phi^*\omega\|^2
+\frac{\alpha}{2}\ip{\phi^*\omega,\d C}+\frac12\|\d C\|^2
-\frac{1}{2}\|C\|^2\geq 0,
\eeq
so the Derrick scaling constraint becomes an integral inequality once
we place the theory in a finite box. This argument, which could be 
called ``Derrick scaling in a finite box'' is of wide applicability in
classical field theory. Note that it does not really require
$B\subset\R^n$ to be a box: any star-shaped domain will do ($B\subset\R^n$
is star-shaped if there exists $x_o\in B$ such that for all $x\in B$
the line segment from $x_o$ to $x$ lies in $B$). 
Of course, for the purposes of numerics,
one would hope to choose the box size $L$ to be
so large (compared with the soliton core radius) that the numerical solution
satisfies (\ref{DC1}) to a good approximation. We will discuss this issue
in more detail in section \ref{numerics}.

\section{Numerical results}\label{numerics}

We used a simple foward differencing scheme to discretize the energy 
functional $E^{(\alpha)}$ on a cubic lattice of spacing $h$ with $N^3$ points,
to give a lattice approximant $E^{(\alpha)}_0:(S^2\times\R^3)^{N^3}\ra\R$.
Given an initial configuration $(\phi,C)$, we used the 
quasi-Newton BFGS
method, as implemented by the TAO and PETSc parallel
numerical libraries
\cite{tao-user-ref} and PETSc\cite{petsc-web-page, petsc-user-ref,
  petsc-efficient},
to find a local minimum of 
$E^{(\alpha)}_0$. We considered the scheme to have converged to a minimum if
the sup norm of the gradient of $E^{(\alpha)}_0$ was less than $0.01h^3$. We
tested the accuracy of this scheme by comparing its results at $\alpha=0$,
for Hopf charges $Q=1,2,3$,
with previous studies of the pure FS model \cite{Sutcliffe:2007ui,Hietarinta:2000ci}
and found good agreement (the
energy minimizers had the same shape, energy and core length to within a few, 
typically 2, percent). By experimenting in the $\alpha=0$ case we found that
a good balance of accuracy and computational speed was obtained with the 
choices $N=480$ and $h=0.0125$, and these values were used for the remaining
calculations.

\begin{figure}
\begin{center}
  \subfloat[][$\alpha=0$]{\includegraphics[width=6cm]{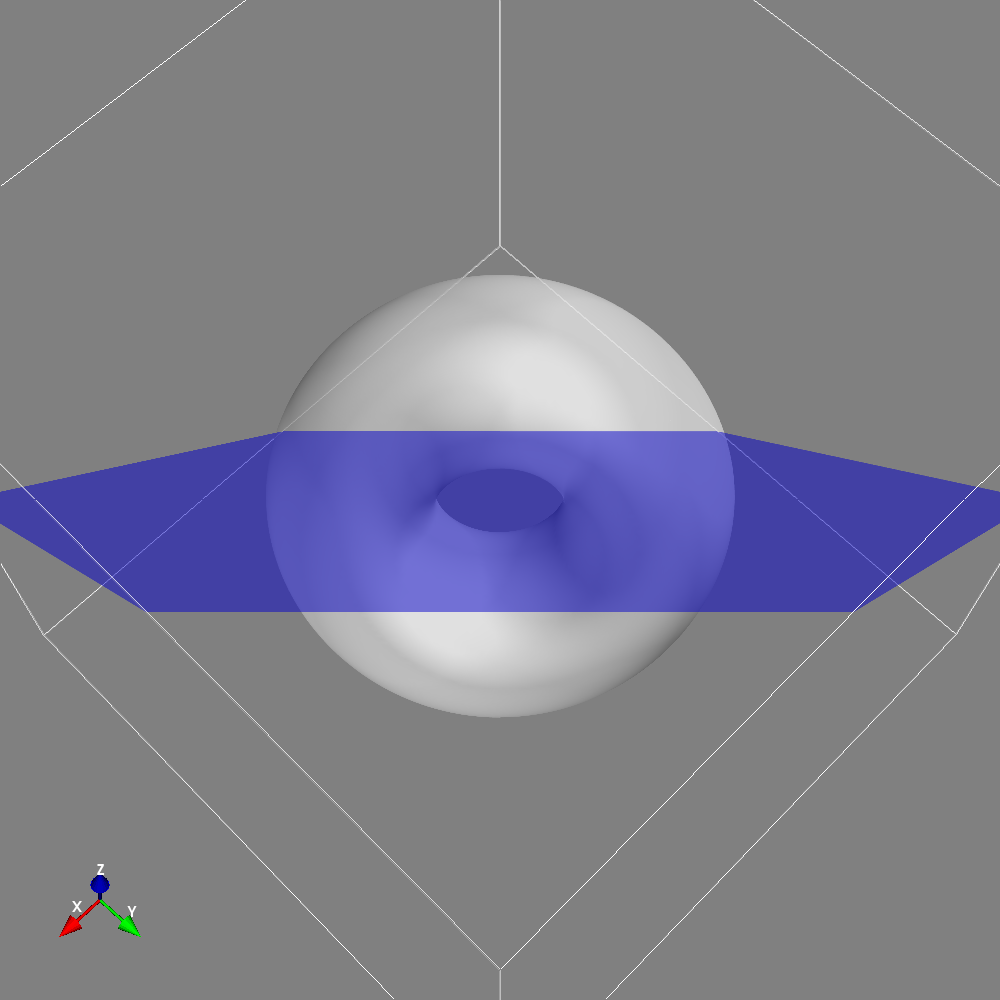}} \hspace*{1.5cm}
  \subfloat[][$\alpha=0.7$]{\includegraphics[width=6cm]{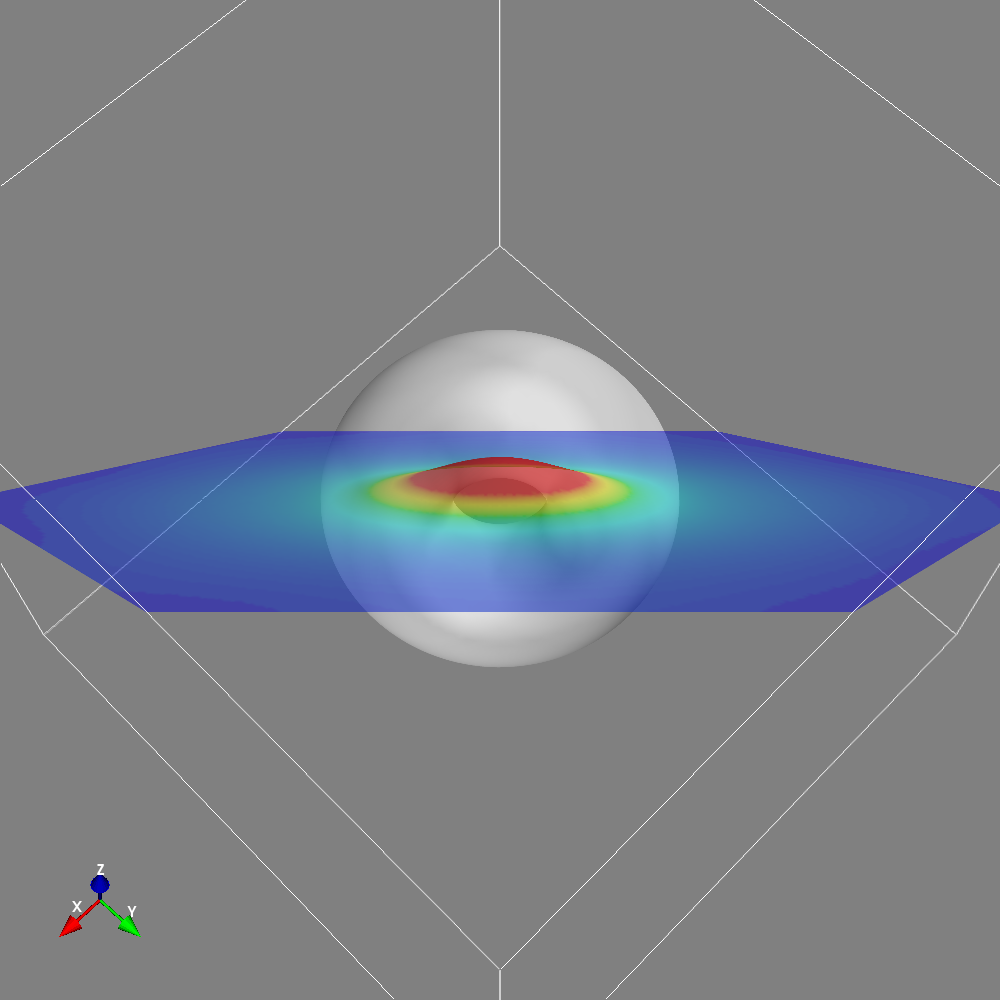}}
\\
  \subfloat[][$\alpha=0.94$]{\includegraphics[width=6cm]{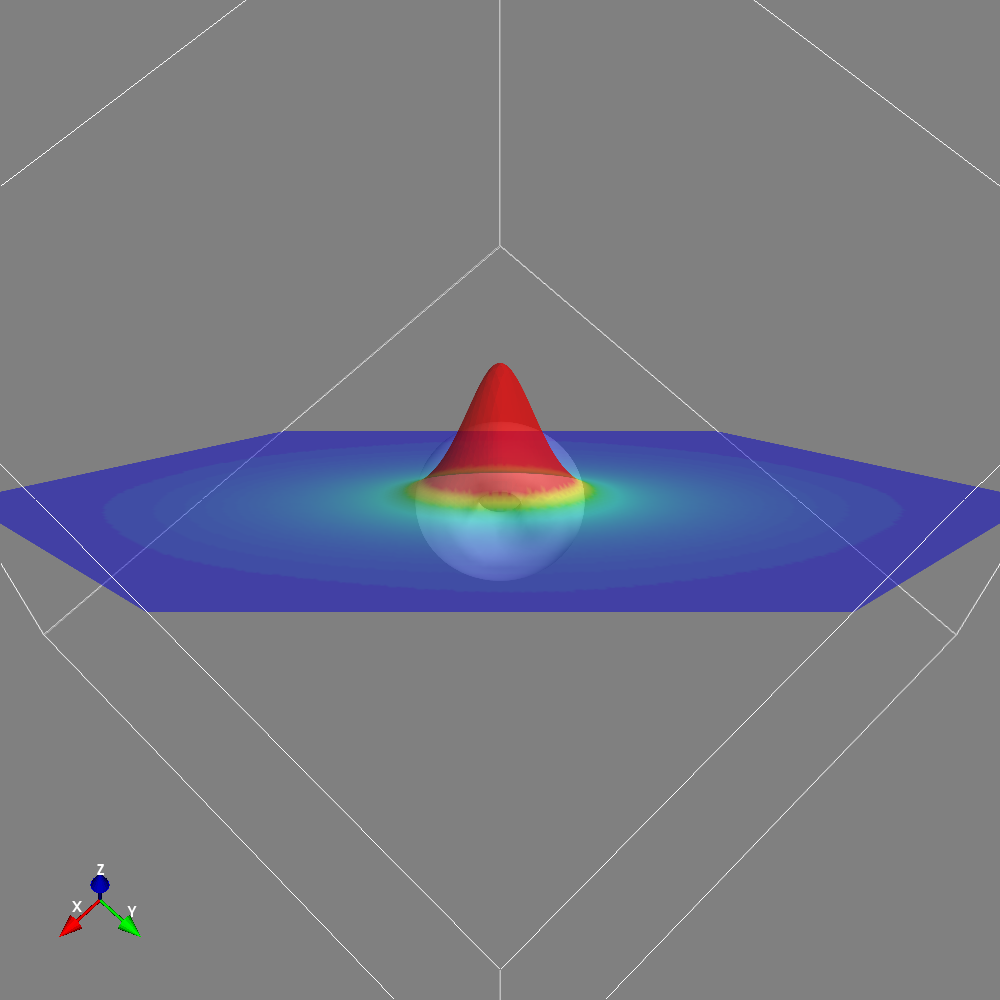}} \hspace*{1.5cm}
  \subfloat[][$\alpha=0.98$]{\includegraphics[width=6cm]{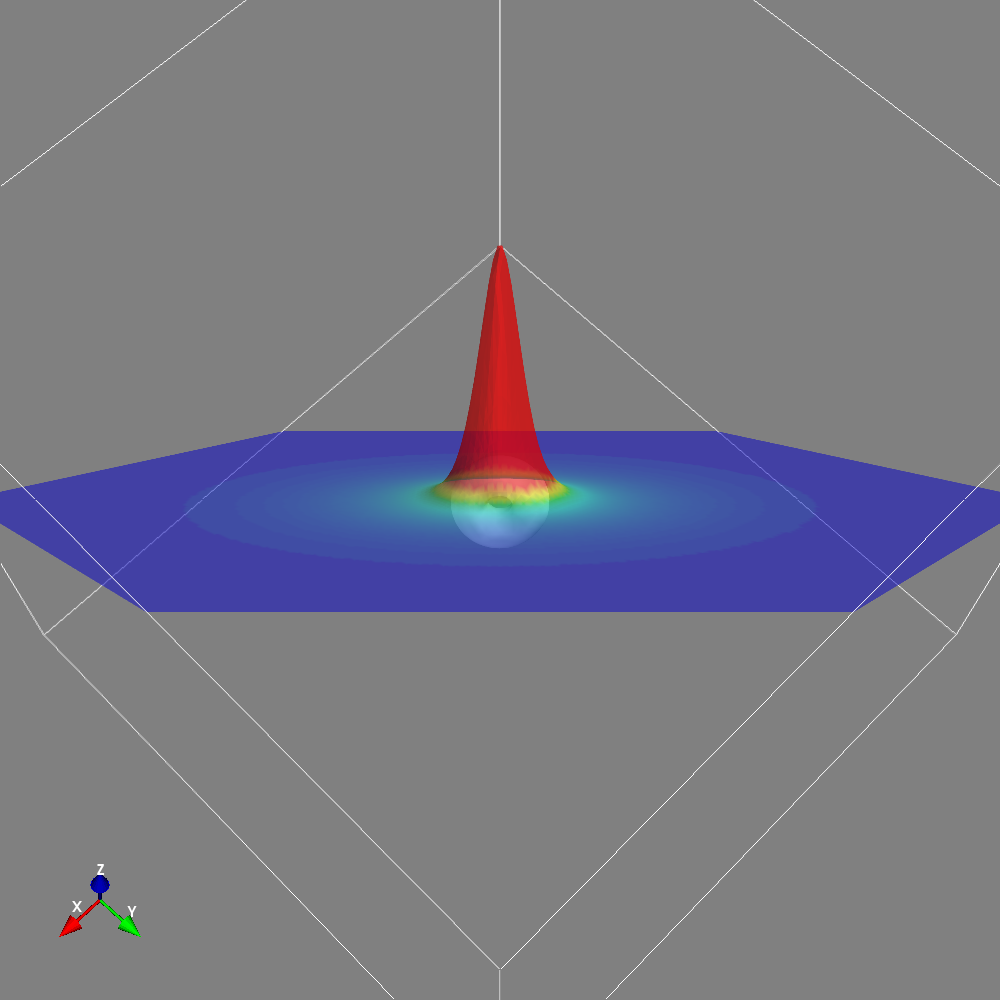}}
\\
  \subfloat[][$\alpha=0.99$]{\includegraphics[width=6cm]{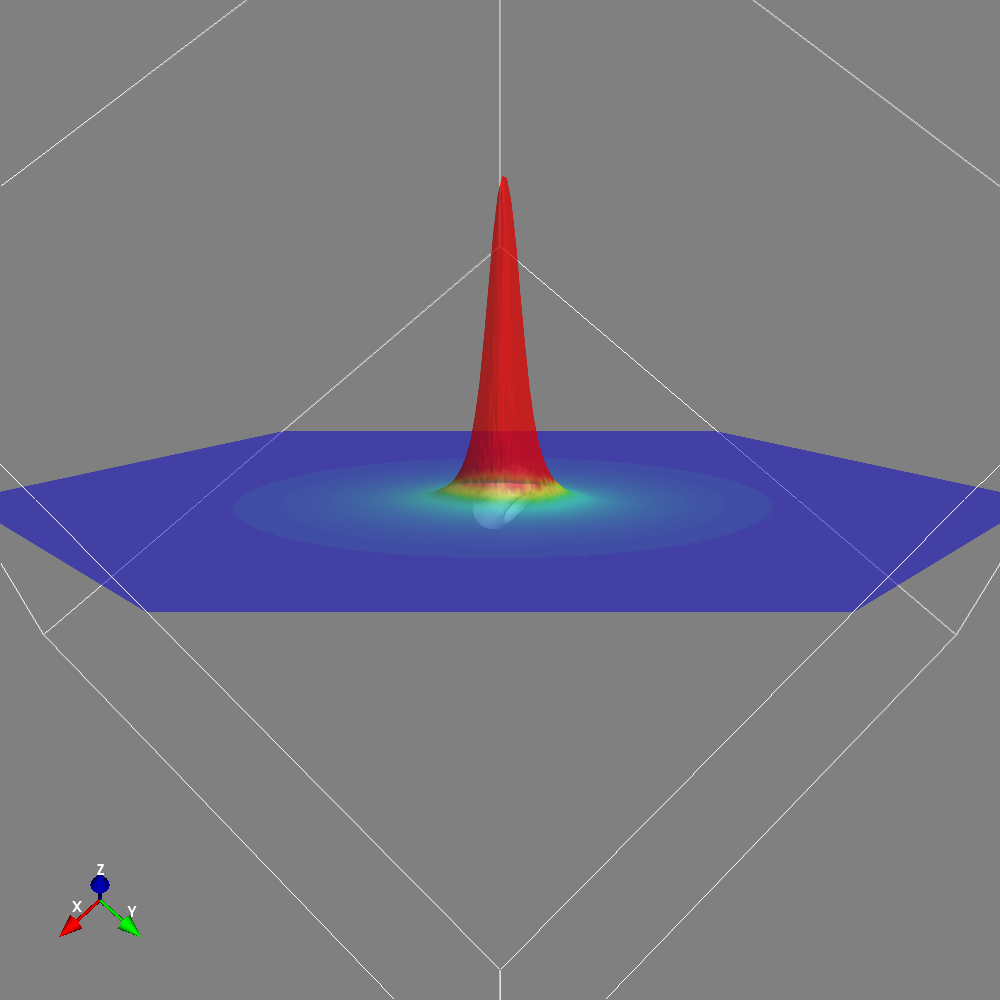}} \hspace*{1.5cm}
  \subfloat[][$\alpha=1$]{\includegraphics[width=6cm]{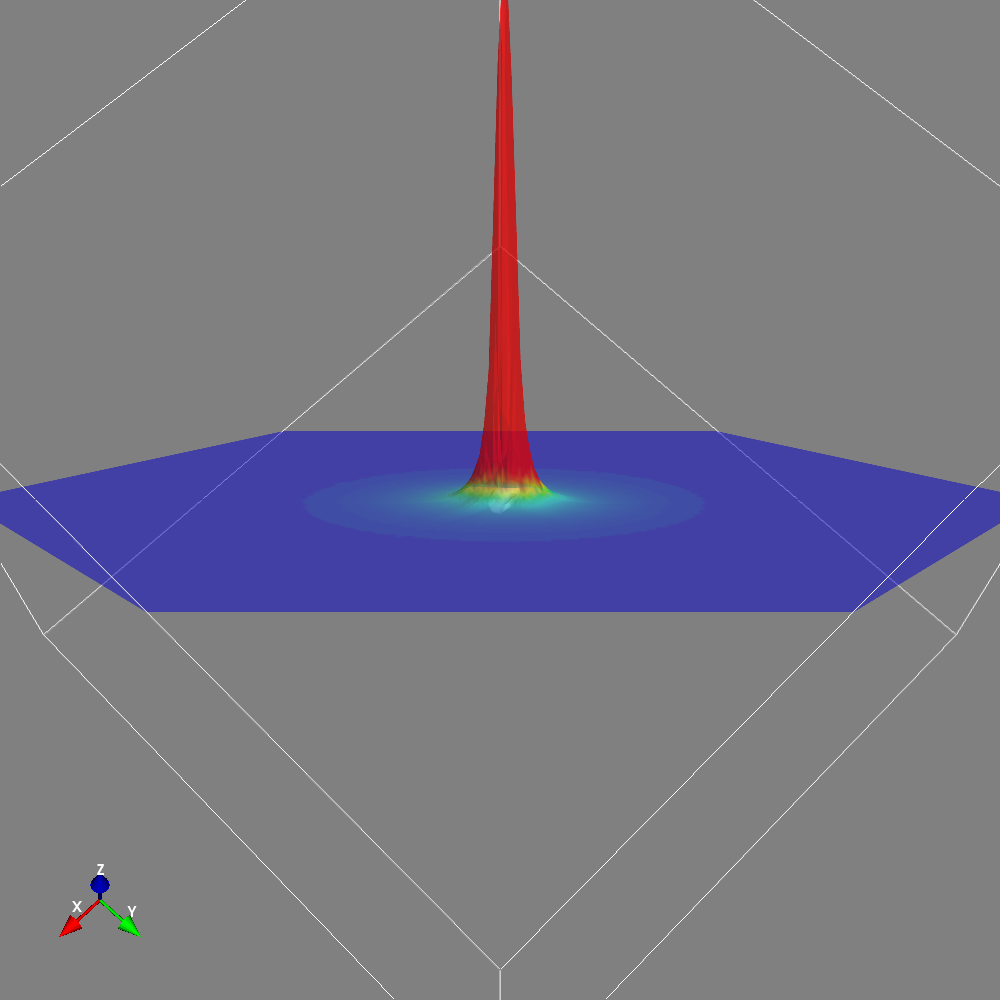}}

\end{center}
\caption{\it Minimizers of $E^{(\alpha)}$ for an increasing
sequence of values of $\alpha$, with Hopf charge $Q=1$. 
The translucent white
surface is the isosurface $\phi_3=0$,
which can be interpreted as bounding the soliton core. The coloured surface
is a plot of $|C|$ on a transverse slice through the soliton core.
Note that as $\alpha\ra 1$, the core shrinks and the $C$ field accumulates
in a singular spike.}\label{shrinking}
\end{figure}

\begin{figure}
\begin{center}
\includegraphics[width=10cm]{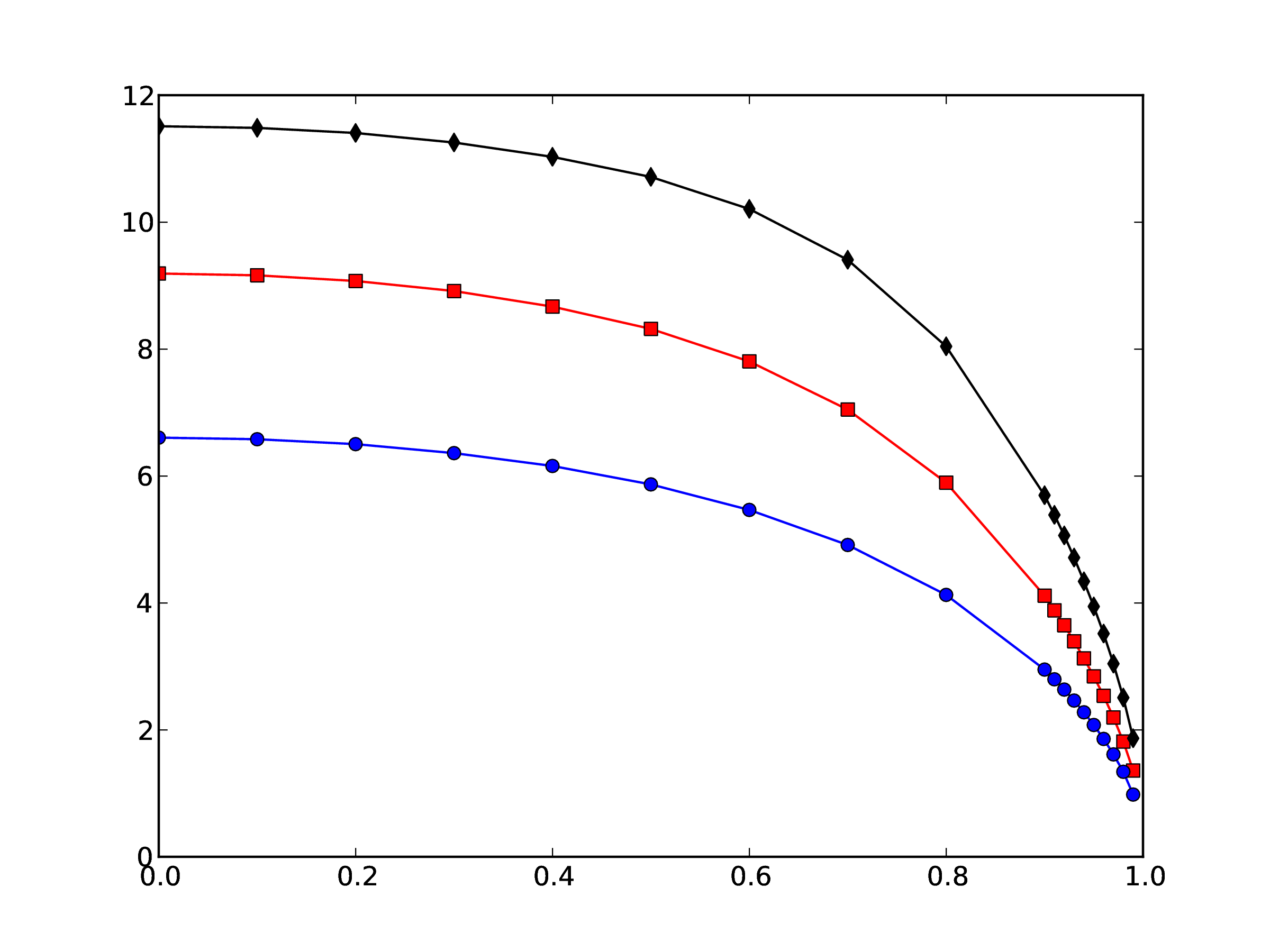}
\end{center}
\caption{\it Core lengths of the $Q=1$ (blue disks), $Q=2$ (red squares), and $Q=3$ (black
  diamonds) minimizers as a function of $\alpha$.}
\label{corelength}
\end{figure}

As described in section \ref{intro}, having found a minimizer of 
$E^{(\alpha)}_0$,
we incremented $\alpha$ slightly, $\alpha'=\alpha+\delta\alpha$,
 and minimized $E^{(\alpha')}_0$ starting with the $\alpha$ minimizer as
our initial guess. In this way, we constructed a curve of minimizers of
$E^{(\alpha)}_0$, parametrized by $\alpha\in[0,1]$, starting at $\alpha=0$
and working towards $\alpha=1$, in each of the homotopy classes $Q=1,2,3$.
In every case, the knot soliton at $\alpha=0$ shrinks rapidly as $\alpha$
approaches $1$, so that, at $\alpha=1$ its core size is comparable to
the lattice spacing. This extremely coarse configuration can 
confidently be identified as a discretization artifact, rather than a
genuine minimizer of $E^{(1)}$, as we shall argue below. The shrinking
process is illustrated in figure \ref{shrinking}, which shows a sequence
of minimizers for increasing $\alpha$ in the case $Q=1$. The results for
$Q=2,3$ look very similar. A quantitative measure of the shrinking is
given in figure \ref{corelength}, which shows the core length of the minimizer
as a function of $\alpha$. Recall that the core is by definition the
preimage curve of $-\phi_\infty$ which, in the cases $Q=1,2,3$ is
a single closed curve. Numerically, we construct this curve
using interpolation tools of the MayaVi Data Visualizer
\cite{ramachandran2010mayavi}. This works well until $\alpha$ is very close to $1$, when
the core structure becomes too small for the lattice to properly resolve, and
the interpolation tool produces a disconnected collection of segments instead
of a closed curve. From this point, the core length calculation is
unreliable, which is why the curves in figure \ref{corelength} stop
slightly short of $\alpha=1$. Clearly, the data up to this point are 
consistent with the core length shrinking rapidly to $0$ as $\alpha\ra 1$.

\begin{figure}
\begin{center}
\includegraphics[width=10cm]{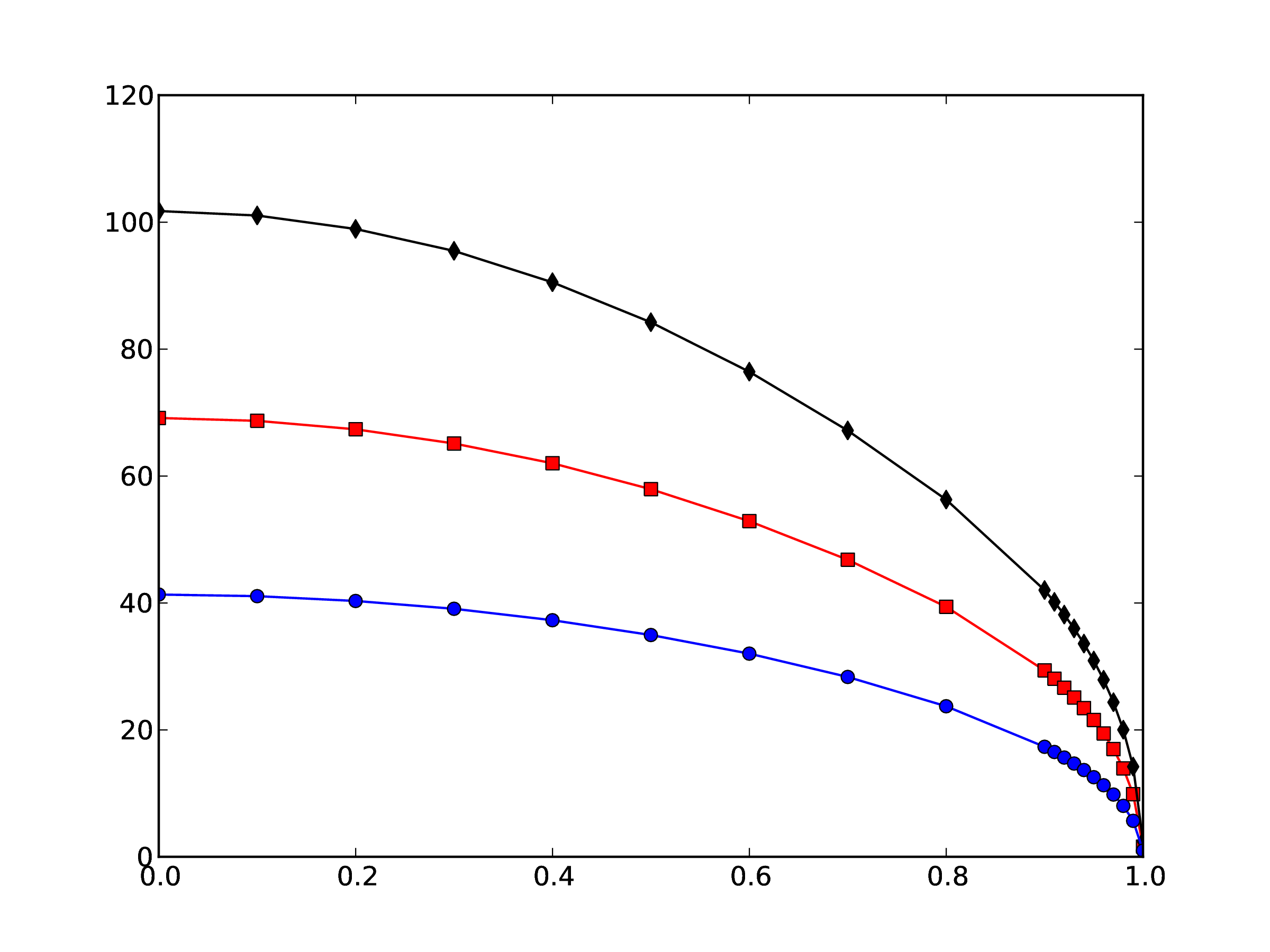}
\end{center}
\caption{\it The squared $L^2$ norm of $\d C+\frac12\phi^*\omega$ of $Q=1$ (blue disks),
  $Q=2$ (red squares),and $Q=3$ (black diamonds) minimizers as a function of $\alpha$.}
\label{norms}
\end{figure}

\begin{figure}
\begin{center}
\includegraphics[width=10cm]{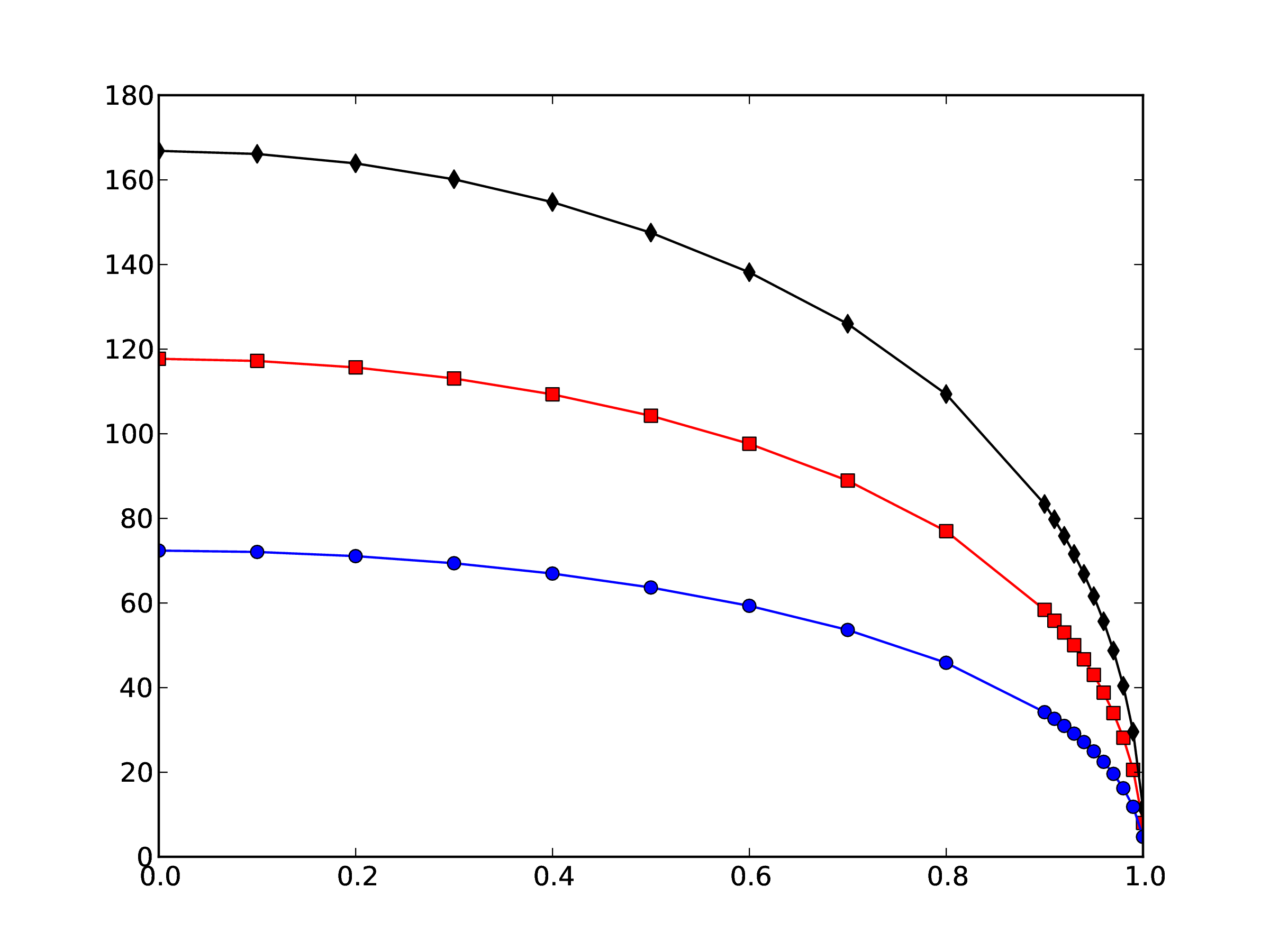}
\end{center}
\caption{\it The energy $E^{(\alpha)}$ of the $Q=1$ (blue disks), $Q=2$ (red squares), and
  $Q=3$ (black diamonds) minimizers as a function of $\alpha$.}
\label{energy}
\end{figure}

Recall we have an energy bound (\ref{ebound}) which vanishes at $\alpha=1$,
and that $E^{(1)}$ has infimum $0$ in every homotopy class, meaning that
every class contains fields of arbitrarily low $E^{(1)}$. Such fields are
constructed by Derrick shrinking among the set of fields satisfying
(\ref{nasty}). Conversely, if (\ref{nasty}) is not satisfied,
the field is protected against shrinking by the presence of
a quartic term in $E^{(1)}$. So if our claim is correct,
that the minimizers shrink and vanish as $\alpha\ra 1$, then they
should satisfy (\ref{nasty}) to a closer and closer approximation as
$\alpha\ra 1$. In figure \ref{norms} we present a graph of
$\|\d C+\frac12\phi^*\omega\|^2$ as a function of $\alpha$ which confirms
that this is indeed what happens: the minimizer loses stability against Derrick 
scaling as $\alpha\ra 1$, and consequently, its energy vanishes rapidly
as $\alpha\ra 1$, as shown in figure \ref{energy}. Note that these
curves are consistent with the bound (\ref{ebound}). 

\begin{figure}
\begin{center}
\includegraphics[width=10cm]{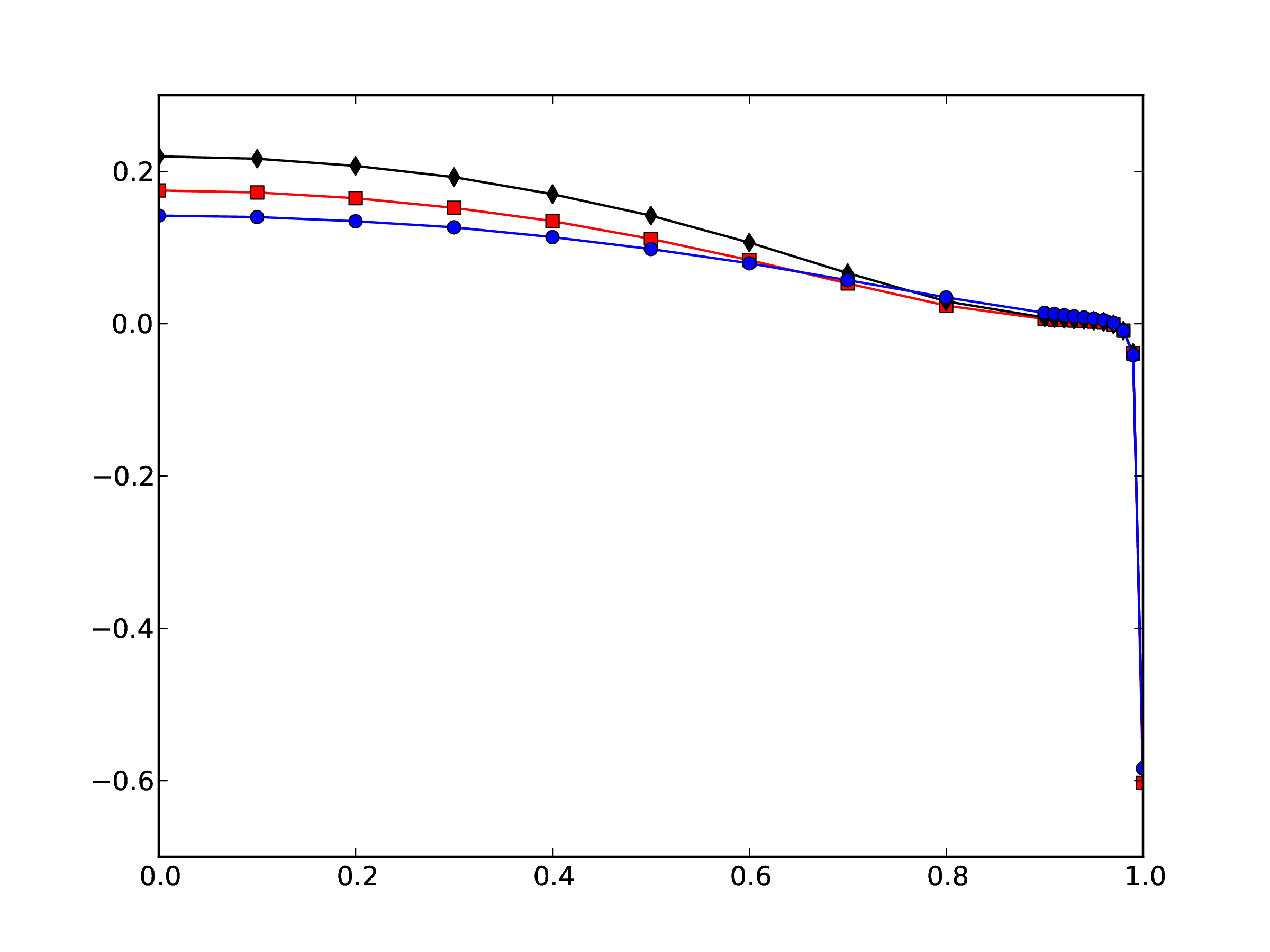}
\end{center}
\caption{\it The Derrick constraint function $D(\alpha)$ of $Q=1$ (blue disks), $Q=2$ (red
  squares), and $Q=3$ (black diamonds) minimizers. This quantity should be non-negative
  for a genuine minimizer in a finite box.}
\label{derrick}
\end{figure}

Clearly the energy
does not vanish exactly at $\alpha=1$, but this is a discretization artifact.
To see this, we define for each $\alpha\in[0,1]$ the quantity
\beq
D(\alpha)=\frac{1}{E^{(\alpha)}(\phi,C)}\left\{-\frac{1}{8}\|\d\phi\|^2+\frac18\|\phi^*\omega\|^2
+\frac{\alpha}{2}\ip{\phi^*\omega,\d C}+\frac12\|\d C\|^2
-\frac{1}{2}\|C\|^2\right\}
\eeq
where $(\phi,C)$ is the minimizer of $E^{(\alpha)}$. This is the left hand
side of the Derrick constraint (\ref{DC1}),
normalized by the total energy
of the field. Hence, as shown in section \ref{ebds}, 
for a minimizer in $\R^3$, one should have $D(\alpha)=0$,
while for a minimizer in a finite box $D(\alpha)\geq 0$.
Figure \ref{derrick} presents a plot of $D(\alpha)$ for our numerical 
minimizers. One sees that $D(\alpha)$
is moderately small  for small $\alpha$, indicating that the boundary of
the finite
box exerts a moderate but appreciable influence on the numerical solution.
As $\alpha$ increases $D(\alpha)$ decreases to very close to zero, 
because the minimizers shrink, so the finite box size becomes numerically
irrelevant. However, as $\alpha$ gets very close to $1$, $D(\alpha)$
becomes negative, indicating that from this point on the minimizers
are strongly affected by discretization effects: if this were
a continuum system, any field with $D(\alpha)<0$ would be
unstable against skrinking.
In particular, $D(1)$ is large and negative,
suggesting that the $\alpha=1$ ``minimizer'' is a lattice artifact, and
not representative of a genuine local minimizer of $E^{(1)}$.

\subsection*{Acknowledgements} This work was supported by the UK 
Engineering and Physical Sciences Research Council.
The authors wish to thank Egor Babaev for useful discussions on this problem.

\end{document}